\documentclass[traditabstract]{aa}

\usepackage{graphicx}
\usepackage[varg]{txfonts}
\usepackage{natbib}
\bibpunct{(}{)}{;}{a}{}{,}
\usepackage{array}
\usepackage{xspace}
\usepackage{lscape}
\usepackage{url}
\usepackage{arydshln}

\usepackage[hyperfootnotes=false, linktocpage=true, breaklinks=true, colorlinks=true, linkcolor=blue, citecolor=blue, urlcolor=blue]{hyperref}
\usepackage[all]{hypcap}

\newcommand{\kms}{\ensuremath{\mathrm{km\ s^{-1}}}\xspace}

\newcommand{\xmm}{{\it XMM-Newton}\xspace}
\newcommand{\chandra}{{\it Chandra}\xspace}

\newcommand{\astroh}{{\it Astro-H}\xspace}

\newcommand{\ngc}{{NGC~5548}\xspace}

\newcommand{\ovi}{\ion{O}{vi}\xspace}
\newcommand{\ovii}{\ion{O}{vii}\xspace}
\newcommand{\fexxiv}{\ion{Fe}{xxiv}\xspace}
\newcommand{\fexxv}{\ion{Fe}{xxv}\xspace}

\newcommand{\sigmav}{\ensuremath{\sigma_{\rm v}}\xspace}

\mathchardef\mhyphen="2D

\begin{document}

\title{Line absorption of He-like triplet lines by Li-like ions}

\subtitle{Caveats of using line ratios of triplets for plasma diagnostics}

\author{
M. Mehdipour \inst{1}
\and
J.S. Kaastra \inst{1,2,3}
\and 
A.J.J. Raassen \inst{1,4}
}

\institute{
SRON Netherlands Institute for Space Research, Sorbonnelaan 2, 3584 CA Utrecht, the Netherlands\\ \email{M.Mehdipour@sron.nl}
\and
Department of Physics and Astronomy, Universiteit Utrecht, P.O. Box 80000, 3508 TA Utrecht, the Netherlands
\and
Leiden Observatory, Leiden University, PO Box 9513, 2300 RA Leiden, the Netherlands
\and
Astronomical Institute ``Anton Pannekoek'', Science Park 904, 1098 XH Amsterdam, University of Amsterdam, the Netherlands
}

\date{Received 15 April 2015 / Accepted 22 May 2015}

\abstract
{
He-like ions produce distinctive series of triplet lines under various astrophysical conditions. However, this emission can be affected by line absorption from Li-like ions in the same medium. We investigate this absorption of He-like triplets and present the implications for diagnostics of plasmas in photoionisation equilibrium using the line ratios of the triplets. Our computations were carried out for the \ovi and \fexxiv absorption of the \ovii and \fexxv triplet emission lines, respectively. The fluorescent emission by the Li-like ions and continuum absorption of the He-like ion triplet lines are also investigated. We determine the absorption of the triplet lines as a function of Li-like ion column density and velocity dispersion of the emitting and absorbing medium. We find \ovi line absorption can significantly alter the \ovii triplet line ratios in optically-thin plasmas, by primarily absorbing the intercombination lines, and to a lesser extent, the forbidden line. Because of intrinsic line absorption by O VI inside a photoionised plasma, the predicted ratio of forbidden to intercombination line intensity for the O VII triplet increases from 4 up to an upper limit of 16. This process can explain the triplet line ratios that are higher than expected and that are seen in some X-ray observations of photoionised plasmas. For the \fexxv triplet, line absorption by \fexxiv becomes less apparent owing to significant fluorescent emission by \fexxiv. Without taking the associated Li-like ion line absorption into account, the density diagnosis of photoionised plasmas using the observed line ratios of the He-like ion triplet emission lines can be unreliable, especially for low-$Z$ ions.
}

\authorrunning{M. Mehdipour et al.}
\titlerunning{Absorption of He-like ion triplets by Li-like ion lines}
\maketitle

\section{Introduction}

Atomic transitions and their resulting spectral lines provide a wealth of useful information about astrophysical plasmas in the universe. High-resolution X-ray spectroscopy is key for any diagnosis of hot plasmas. Over the past 15 years, the grating spectrometers onboard \chandra (LETGS, \citealt{Brink00} and HETGS, \citealt{Caniz00}) and \xmm (RGS, \citealt{denH01}) have been instrumental in extracting important physical information from hot plasmas. One example of important spectral lines is that of He-like ions, which have been used for density and temperature diagnostics. 

The first six excited states of a He-like atom have slightly different energy levels, depending on the angular momentum and spin of the excited electron, and have four line transitions to the ground state: the resonance line $w$ (${1{s}\, 2{p}\ ^{1}{\rm{P}}_{1} \to 1{s}^{2}\ ^{1}{\rm{S}}_{0}}$), the intercombination lines $x$ (${1{s}\, 2{p}\ ^{3}{\rm{P}}_{2} \to 1{s}^{2}\ ^{1}{\rm{S}}_{0}}$) and $y$ (${1{s}\, 2{p}\ ^{3}{\rm{P}}_{1} \to 1{s}^{2}\ ^{1}{\rm{S}}_{0}}$), and the forbidden line $z$ (${1{s}\, 2{s}\ ^{3}{\rm{S}}_{1} \to 1{s}^{2}\ ^{1}{\rm{S}}_{0}}$). Counting the $x$ and $y$ lines as one line, they are often called the He-like triplet. For the analysis of solar coronae, \citet{Gabr69} first proposed that the relative intensities of the lines in a He-like triplet can be used for temperature and density diagnostics. They introduced line ratios, which depend on electron density $n_{\rm{e}}$, and temperature $T_{\rm{e}}$, which are ${R\,({n_{\rm{e}}}) = z\, /\, (x + y)}$ and ${G\,({T_{\rm{e}}}) = (z + (x + y))\, /\, w}$, where $w$, $x+y$, and $z$ stand for the intensities of the resonance, intercombination and forbidden lines, respectively. Since then other studies have advanced our understanding of He-like ion diagnostics in collisional ionisation equilibrium (CIE) plasmas, as well as extending the diagnosis to plasmas in photoionisation equilibrium (PIE) and non-equilibrium ionisation (NEI). For example, \citet{Por00} have computed and compared the $R$ and $G$ ratios for six He-like ions for plasmas under CIE, PIE, and hybrid conditions. For a review of He-like diagnostic studies, see \citet{Porq10} and references therein.

The absorption lines of a Li-like ion occur close to the emission lines of its He-like ion triplet. The absorption lines can therefore affect the intensity of the triplet lines, in particular, those of the intercombination lines as shown in the present study. Without taking the Li-like ion line absorption intrinsic to a medium emitting He-like triplets into account, incorrect triplet line ratios can be obtained, which result in misleading plasma diagnostics. 

%
\begin{table*}[!]
\begin{minipage}[t]{\hsize}
\setlength{\extrarowheight}{3pt}
\caption{Atomic parameters of those \ovi and \fexxiv transitions, which contribute to the line absorption of the \ovii and \fexxv triplet lines.}
\label{lines_table}
\centering
\small
\renewcommand{\footnoterule}{}
\begin{tabular}{l c c | l c c c c c}
\hline \hline
Ion & $\lambda_{\rm c}$ ($\AA$) & $E_{\rm c}$ (keV) & Upper energy level & $f_{\rm{osc}}$ & $A_{\rm{rad}}$ (s$^{-1}$) & $A_{\rm{aut}}$ (s$^{-1}$)  & $A_{\rm{tot}}$ (s$^{-1}$) & $\omega$  \\
\hline

\ion{O}{vi}  & $21.7881$ & $0.569046$ & 	$1s2s(^1{\rm S})2p\ ^2{\rm P}^{\rm }_{3/2}$  	& $5.27 \times 10^{-2}$   		& $3.68 \times 10^{11}$			& $6.98 \times 10^{13}$		& $7.02 \times 10^{13}$ & $5.24 \times 10^{-3}$ \\
\ion{O}{vi}  & $21.7891$ & $0.569020$ & 	$1s2s(^1{\rm S})2p\ ^2{\rm P}^{\rm }_{1/2}$ 	& $2.82 \times 10^{-2}$ 		& $3.93 \times 10^{11}$			& $6.96 \times 10^{13}$ 		& $7.00 \times 10^{13}$ & $5.61 \times 10^{-3}$ \\
\ion{O}{vi}  & $22.0189$ & $0.563081$ & 	$1s2s(^3{\rm S})2p\ ^2{\rm P}^{\rm }_{3/2}$  	& $3.85 \times 10^{-1}$		& $2.64 \times 10^{12}$			& $8.72 \times 10^{12}$		& $1.14 \times 10^{13}$ & $2.32 \times 10^{-1}$ \\
\ion{O}{vi}  & $22.0205$ & $0.563040$ & 	$1s2s(^3{\rm S})2p\ ^2{\rm P}^{\rm }_{1/2}$  	& $1.91 \times 10^{-1}$		& $2.61 \times 10^{12}$			& $9.57 \times 10^{12}$ 		& $1.22 \times 10^{13}$ & $2.14 \times 10^{-1}$ \\

\hline

\ion{Fe}{xxiv}  & $1.8563$ & $6.6791$ & 	${1s}(^2{\rm S}){2s2p}(^1{\rm P}^{\rm })\ ^2{\rm P}^{\rm }_{3/2}$	& $9.80 \times 10^{-4}$		& $9.51 \times 10^{11}$	& $1.10 \times 10^{14}$ 		& $1.11 \times 10^{14}$ & $8.57 \times 10^{-3}$ \\
\ion{Fe}{xxiv}  & $1.8570$ & $6.6764$ & 	${1s}(^2{\rm S}){2s2p}(^1{\rm P}^{\rm })\ ^2{\rm P}^{\rm }_{1/2}$	& $9.19 \times 10^{-2}$ 		& $1.77 \times 10^{14}$	& $8.18 \times 10^{13}$ 		& $2.59 \times 10^{14}$ & $6.83 \times 10^{-1}$ \\
\ion{Fe}{xxiv}  & $1.8611$ & $6.6619$ & 	${1s}(^2{\rm S}){2s2p}(^3{\rm P}^{\rm })\ ^2{\rm P}^{\rm }_{3/2}$	& $4.97 \times 10^{-1}$		& $4.79 \times 10^{14}$	& $5.80 \times 10^{8}$  		& $4.79 \times 10^{14}$ & $1.00$ \\
\ion{Fe}{xxiv}  & $1.8635$ & $6.6535$ & 	${1s}(^2{\rm S}){2s2p}(^3{\rm P}^{\rm })\ ^2{\rm P}^{\rm }_{1/2}$	& $1.63 \times 10^{-1}$		& $3.13 \times 10^{14}$	& $3.89 \times 10^{13}$ 		& $3.52 \times 10^{14}$ & $8.89 \times 10^{-1}$ \\
\ion{Fe}{xxiv}  & $1.8738$ & $6.6167$ & 	${1s}(^2{\rm S}){2s2p}(^3{\rm P}^{\rm })\ ^4{\rm P}^{\rm }_{3/2}$	& $1.66 \times 10^{-2}$   	& $1.57 \times 10^{13}$	& $7.23 \times 10^{11}$ 		& $1.64 \times 10^{13}$ & $9.57 \times 10^{-1}$ \\

\hline

\end{tabular}
\end{minipage}

\tablefoot{
$\lambda_{\rm c}$ and $E_{\rm c}$ are the line-centre wavelength and energy, respectively. $A_{\rm{rad}}$, $A_{\rm{aut}}$ and $A_{\rm{tot}}$ are the radiative, autoionisation and total transition probabilities, respectively. The oscillator strength $f_{\rm{osc}}$ and fluorescence yield $\omega$ are dimensionless. For all lines the lower energy level is ${1s^{2} 2s}\ ^2\rm{S}_{1/2}$. 
}

\end{table*}

\section{Diagnostics of photoionised plasmas using X-ray detections of He-like ion triplets}
\label{triplet_obs_sect}

In the literature there are reported X-ray detections and diagnostics of He-like triplets in emission from photoionised plasmas in active galactic nuclei (AGN). \citet{Colli01} identified the forbidden and intercombination lines of both \ovii and \ion{Ne}{ix} in\object{ NGC~4051}. The $R$-ratio was found to be about 5 for both triplets. Using this $R$-ratio and the theoretical calculations of \cite{Por00}, they obtain an upper limit of ${4\times 10^{10}\ {\rm{cm}}^{-3}}$ for the density $n_{\rm{e}}$. \citet{McKe03} have estimated an upper limit of ${R < 3}$ for \ion{Ne}{ix} and ${R < 10}$ for \ion{O}{vii} in \object{NGC~4593}; these constraints imply different densities of ${n_{\rm{e}} < 2 \times 10^{12}\ {\rm{cm}}^{-3}}$ and ${n_{\rm{e}} < 8 \times 10^{10}\ {\rm{cm}}^{-3}}$, respectively. Recently, \citet{Landt15} found that the \ovii $R$-ratio from RGS spectra of \object{NGC 4151} is significantly higher than the one from their theoretical calculations for a PIE plasma using the {\tt Cloudy} code \citep{Fer98}. For three of their observing epochs, in which the flux of the \ovii lines is best constrained, they measure ${R = 6.0 \pm 0.9}$, ${7.2 \pm 1.0}$ and ${11.9 \pm 2.5}$, whereas the predicted theoretical upper limit is ${R = 4.0}$. They also find that an additional contribution from a CIE plasma would not increase the predicted $R$-ratio high enough to reach the observed $R$. Finally, the \ovii $R$-ratio from RGS spectroscopy is observed to be ${5.5 \pm 1.7}$ in \object{Mrk~3} \citep{Poun05} and ${9.3 \pm 4.1}$ in \object{NGC 1068} \citep{Kraem15}. After correcting for absorption using {\tt Cloudy}, \citet{Kraem15} report that the absorption-corrected ${R = 7.7 \pm 2.8}$, which is still higher than their predicted $R$ of 3.9 to 4.0. In general, the line ratios from X-ray detections of He-like triplets in AGN indicate some degree of deviation from theoretical predications for a PIE plasma, which may be due to Li-like ion line absorption as we argue in this paper.

\section{Computation of the Li-like ion line absorption of the He-like ion triplet lines}
\label{compute_sect}

The absorbed line intensity $I(\lambda)$ is related to the initial intensity $I_{0}(\lambda)$ before absorption according to ${I(\lambda) = I_{0}(\lambda)\, T(\lambda)}$, where $T(\lambda)$ is the transmittance of the absorbing medium as a function of wavelength $\lambda$. For the case of a slab containing an emitting and absorbing medium, the transmittance ${T(\lambda) = (1 - \mathrm{e}^{-\tau(\lambda) / \cos \theta})\, /\, (\tau(\lambda) / \cos \theta)}$, where $\tau(\lambda)$ is the optical depth of the medium and $\theta$ is the angle between the line of sight and the normal to the plane of the slab. In this work, we adopt ${\theta = 0}$. The optical depth ${\tau(\lambda) = \tau_{0}\, \phi(\lambda)}$, where $\tau_{0}$ is the optical depth at the line-centre $\lambda_{\rm c}$ and $\phi(\lambda)$ is the line profile. The optical depth ${\tau_{0} = \alpha\, h\, \lambda_{\rm c}\, f_{\rm osc}\, N_{\rm ion}\, /\, 2\sqrt{2\pi}\, m_{\rm e}\, \sigma_{{\rm v}}}$, where $\alpha$ is the fine structure constant, $h$ the Planck constant, $f_{\rm osc}$ the oscillator strength, $N_{\rm ion}$ the column density of the absorbing ion, $m_{\rm{e}}$ the electron mass and $\sigma_{\rm v}$ the velocity dispersion. The line profile $\phi(\lambda)$ is modelled with a Voigt line profile $H(a,y)$, where ${a=A_{\rm tot} \lambda / 4 \pi b}$, ${y=c\, (\lambda - \lambda_{\rm c}) / b \lambda_{\rm{c}}}$. Here, ${b=\sqrt{2} \sigma_{\rm{v}}}$ and $A_{\rm tot}$ is the total transition probability, which is the sum of radiative $A_{\rm rad}$ and autoionisation $A_{\rm rad}$ transition probabilities. The atomic parameters of the \ovi and \fexxiv transitions, which contribute to the line absorption of the \ovii and \fexxv triplet emission lines, are given in Table \ref{lines_table}. The parameters are obtained from the Flexible Atomic Code ({\tt FAC}) \citep{Gu11}. 

The emission lines of the He-like ion triplets were modelled using a Gaussian profile. The Voigt profile was not required as the Lorentzian contribution was found to be minimal for these lines, hence the Gaussian profile provides a decent approximation. The line-centre wavelengths of the \ovii triplet lines are ${21.6020\ \AA}$ ($w$), ${21.8044\ \AA}$ ($x$), ${21.8070\ \AA}$ ($y$) and ${22.1012\ \AA}$ ($z$). The line-centre energies of the \fexxv triplet lines are $6.7001$~keV ($w$), $6.6822$~keV ($x$), $6.6678$~keV ($y$) and $6.6365$~keV ($z$). The wavelengths or energies of the He-like triplets and the Li-like absorption lines are from the NIST Atomic Spectra Database v5 and \citet{Schm04}. In our computations, $\sigma_{\rm v}$ for the He-like ions was fixed to that of the Li-like ions as they are taken to be part of the same medium. 

Following absorption by a Li-like ion, the excited ion de-excites by filling the vacancy in its inner shell with an outer-shell electron. The released energy emerges either as an Auger electron or a fluorescent photon. As the latter results in effectively cancelling out the Li-like line absorption, its contribution needs to be taken into account. The fluorescence yield $\omega$ is given by ${\omega = A_{\rm rad} / (A_{\rm rad} + A_{\rm aut})}$ and generally increases with the nuclear charge of the ion, thus the \fexxiv lines have higher $\omega$ than the \ovi lines (see Table~\ref{lines_table}). The observed effect of the fluorescent emission was approximated in our calculations by multiplying the $\tau$ of each absorption line with ${(1 - \omega)}$. So at ${\omega = 0}$, $\tau$ remains unchanged and at ${\omega = 1}$, $\tau$ becomes zero as all the absorbed line photons are re-emitted as fluorescent photons. 

\begin{figure*}[!]
\centering
\resizebox{0.92\hsize}{!}{\hspace{-1cm}\includegraphics[angle=0]{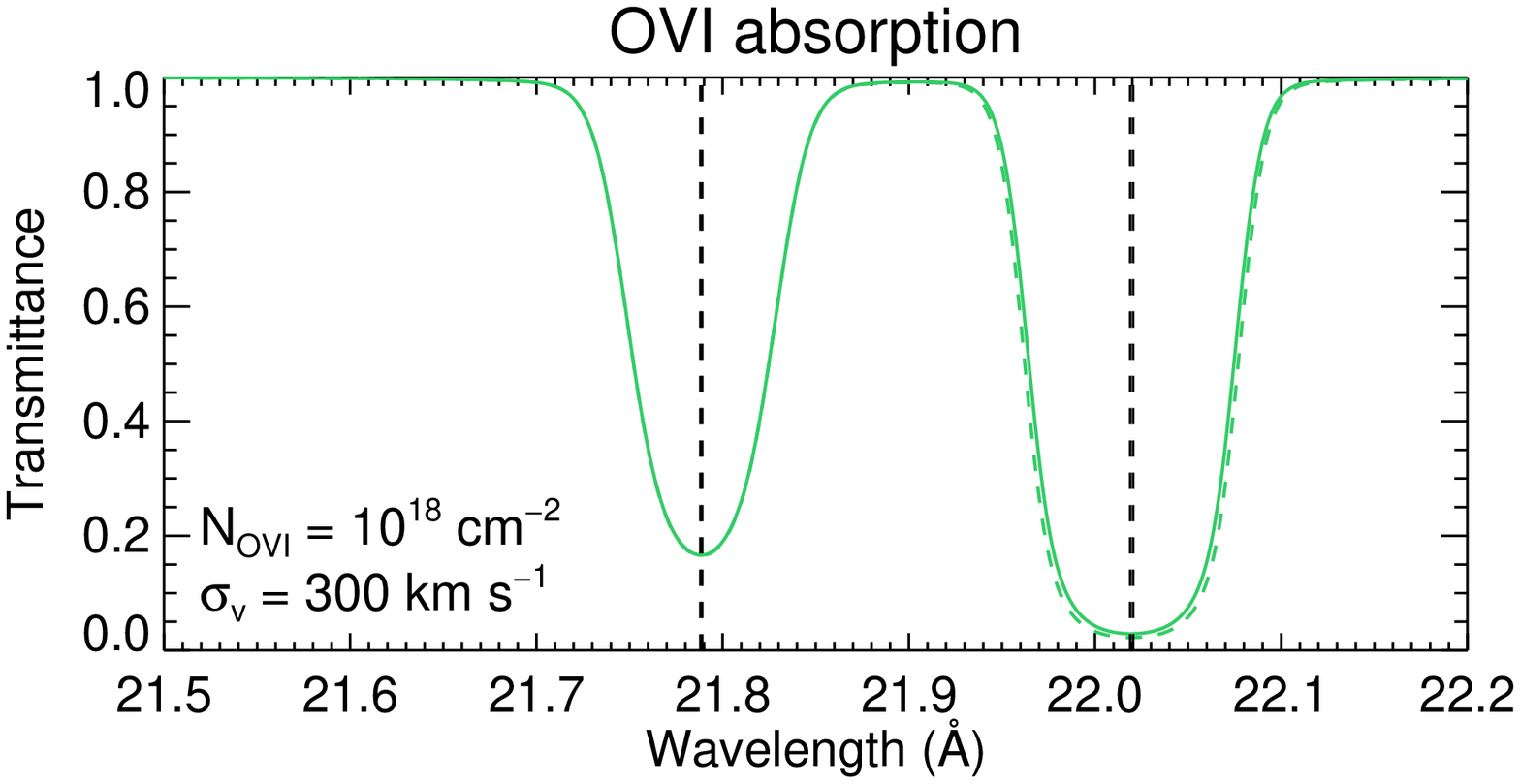}\hspace{-1.4cm}\includegraphics[angle=0]{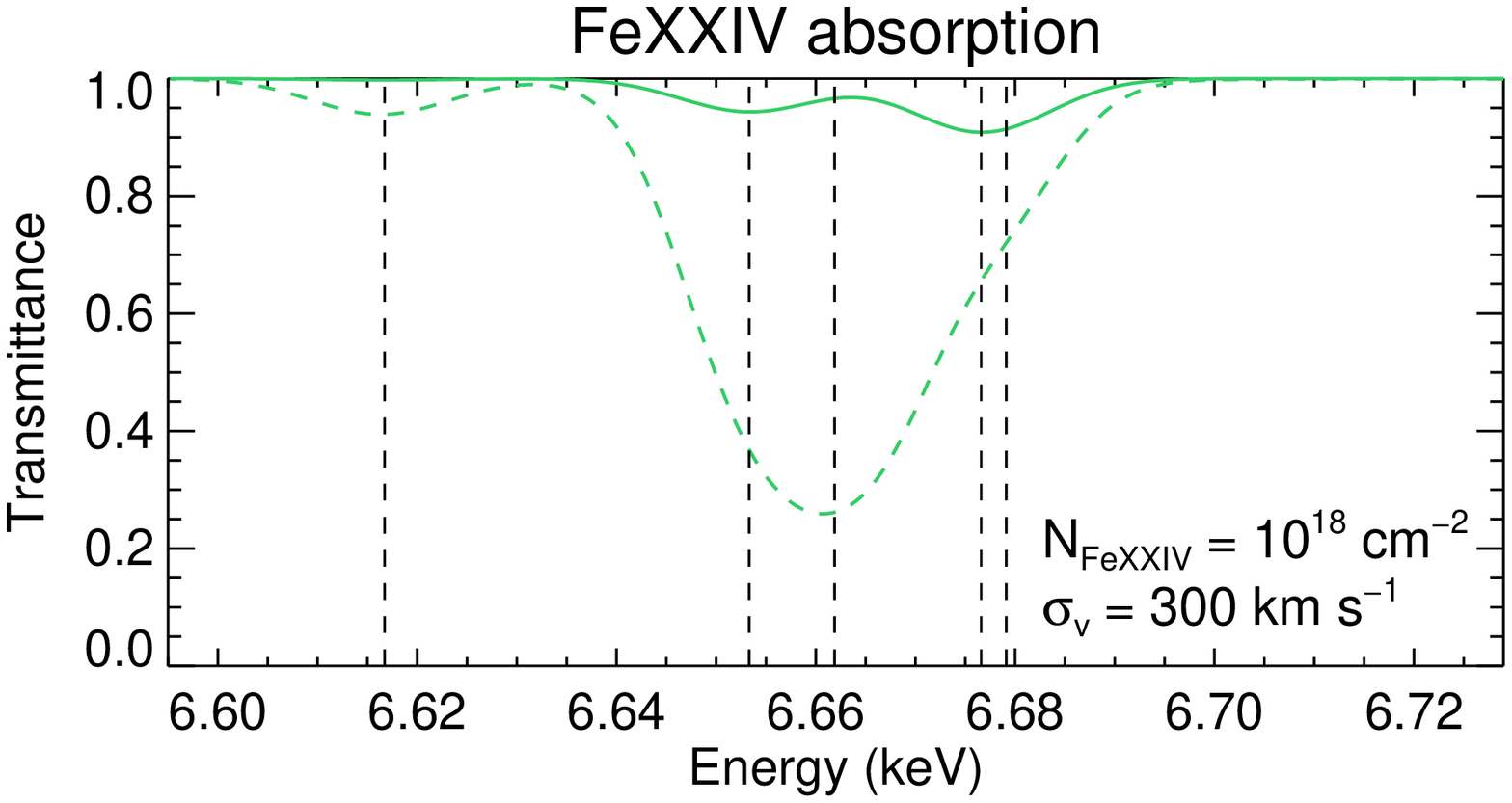}}
\resizebox{0.92\hsize}{!}{\hspace{-1cm}\includegraphics[angle=0]{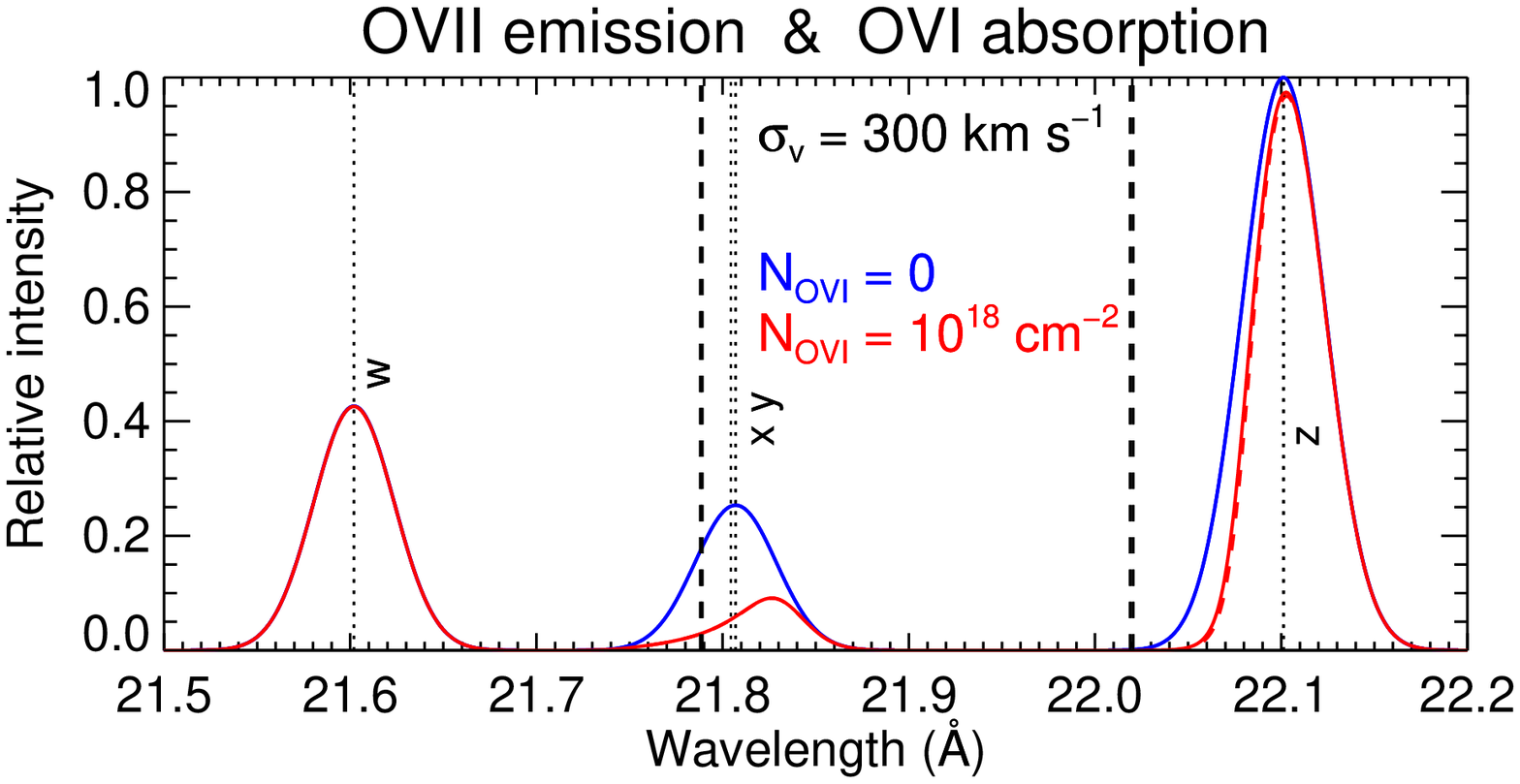}\hspace{-1.4cm}\includegraphics[angle=0]{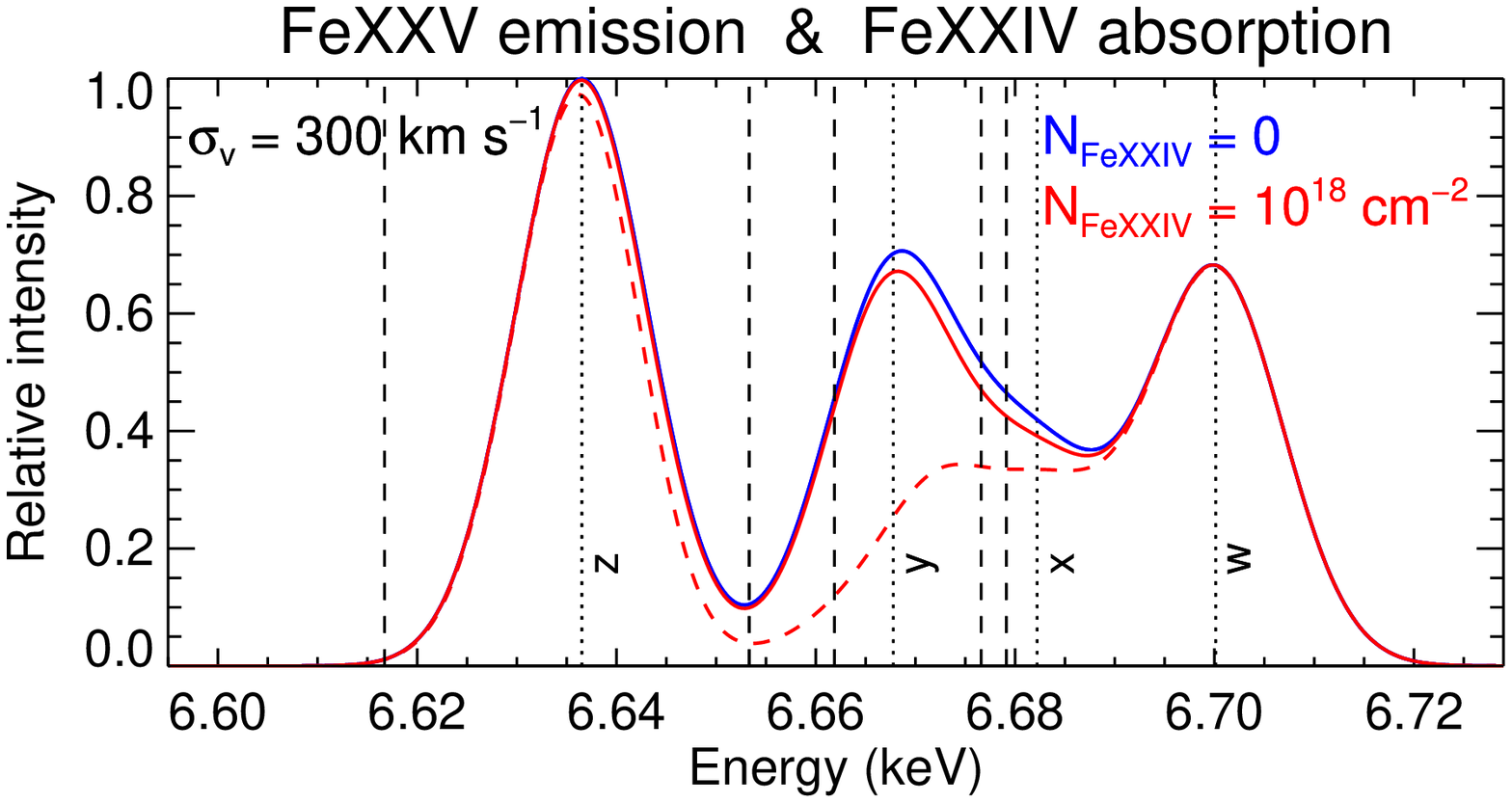}}
\caption{{\it Top panels}: examples of \ovi and \fexxiv transmittance $T(\lambda)$ with ${N_{\rm Li \mhyphen like} = 10^{18}\ {\rm cm}^{-2}}$ and ${\sigmav = 300\ \kms}$. {\it Bottom panels}: examples of \ovii and \fexxv triplet emission lines (for a PIE plasma case) absorbed by \ovi and \fexxiv lines, respectively. In all panels the effect of fluorescent emission is indicated using the dashed curves (excluding fluorescence) compared to the solid curves (including fluorescence). The $\lambda_{\rm c}$ or $E_{\rm c}$ of the He-like (\ovii and \fexxv) emission lines are indicated with vertical dotted lines and those of the Li-like (\ovi and \fexxiv) absorption lines with vertical dashed lines.
\vspace{0.5cm}
}
\label{triplets_fig}
\end{figure*}

\begin{figure*}[!]
\centering
\resizebox{0.94\hsize}{!}{\includegraphics[angle=0]{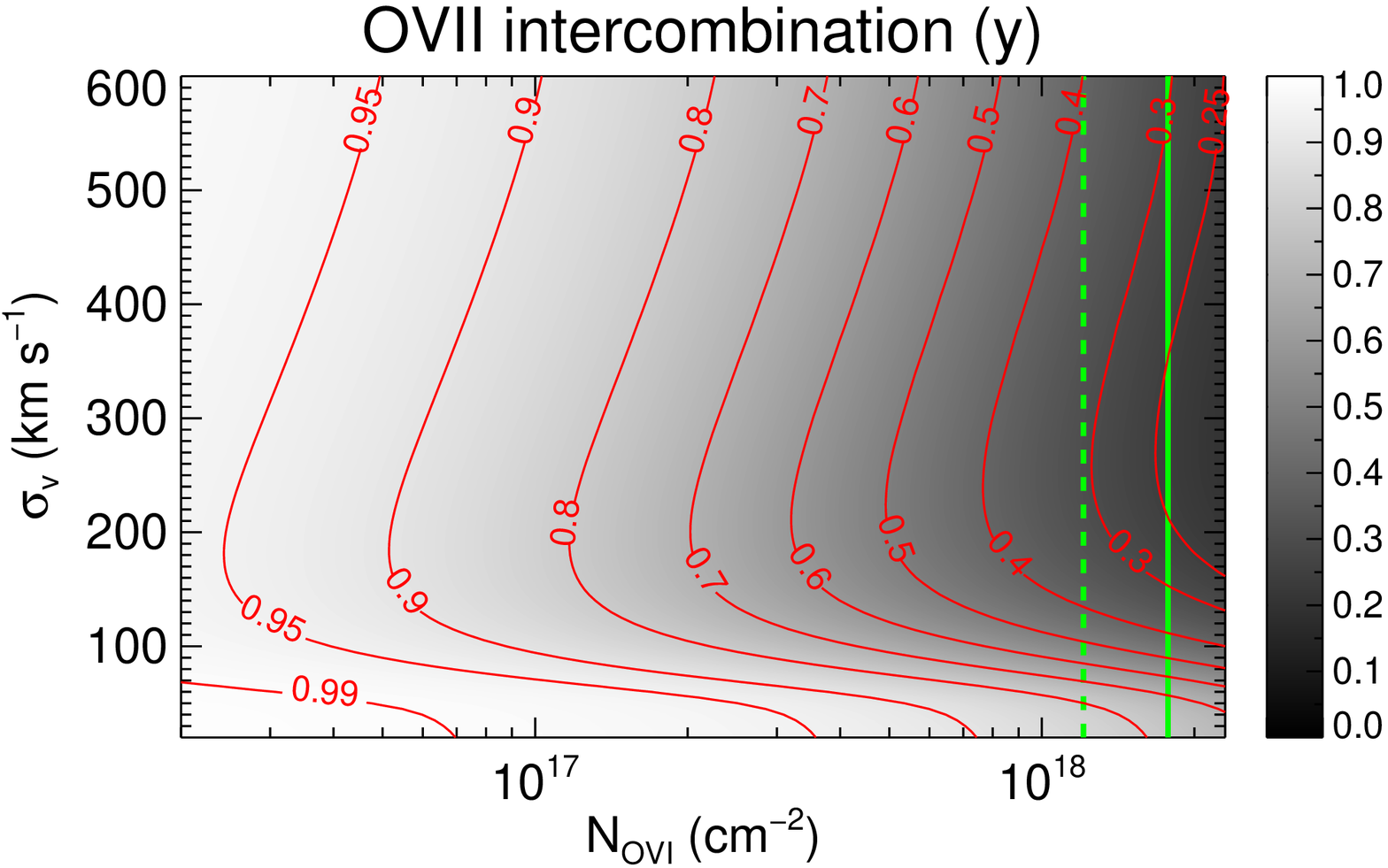}\hspace{1.0cm}\includegraphics[angle=0]{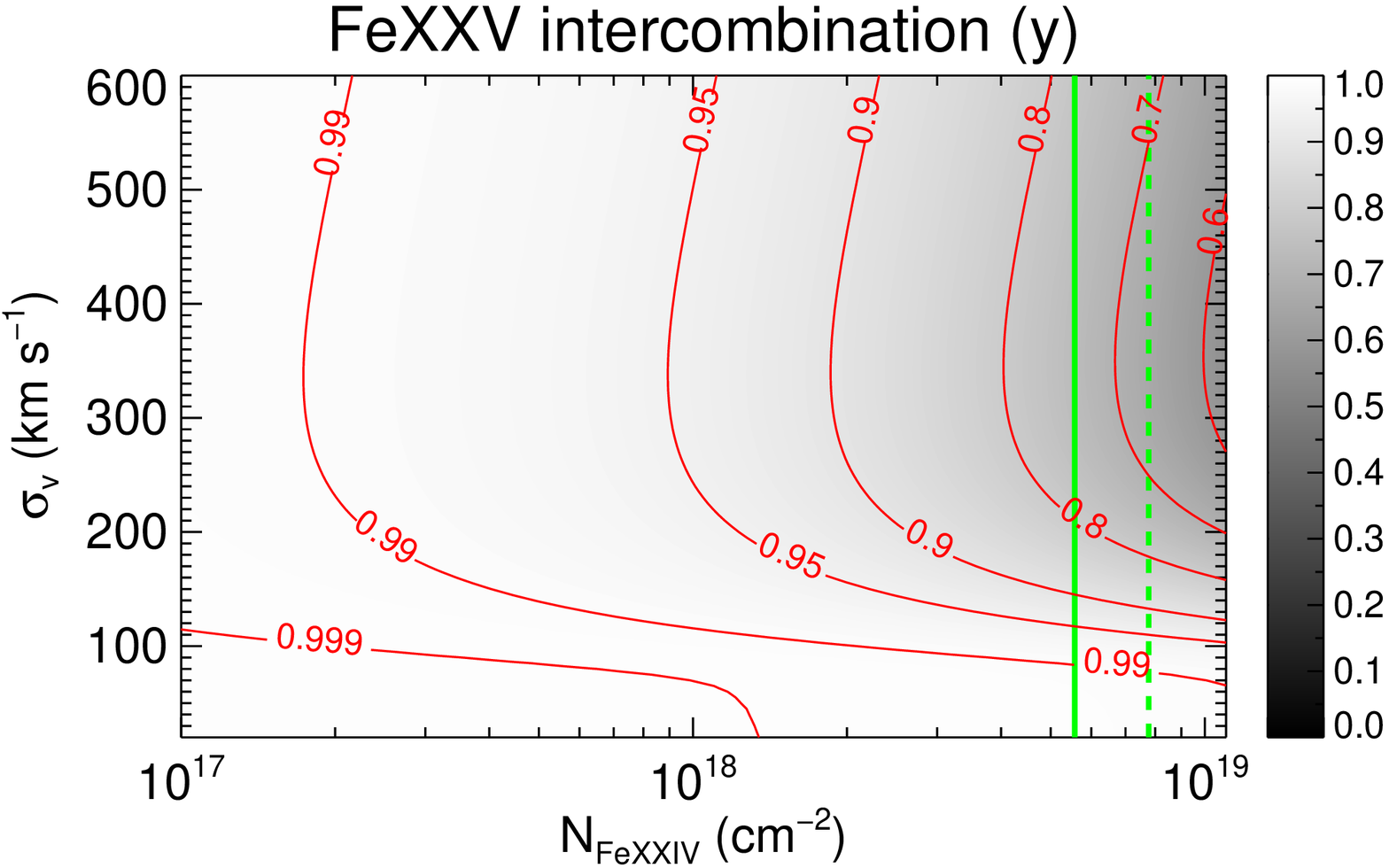}}\vspace{0.2cm}
\resizebox{0.94\hsize}{!}{\includegraphics[angle=0]{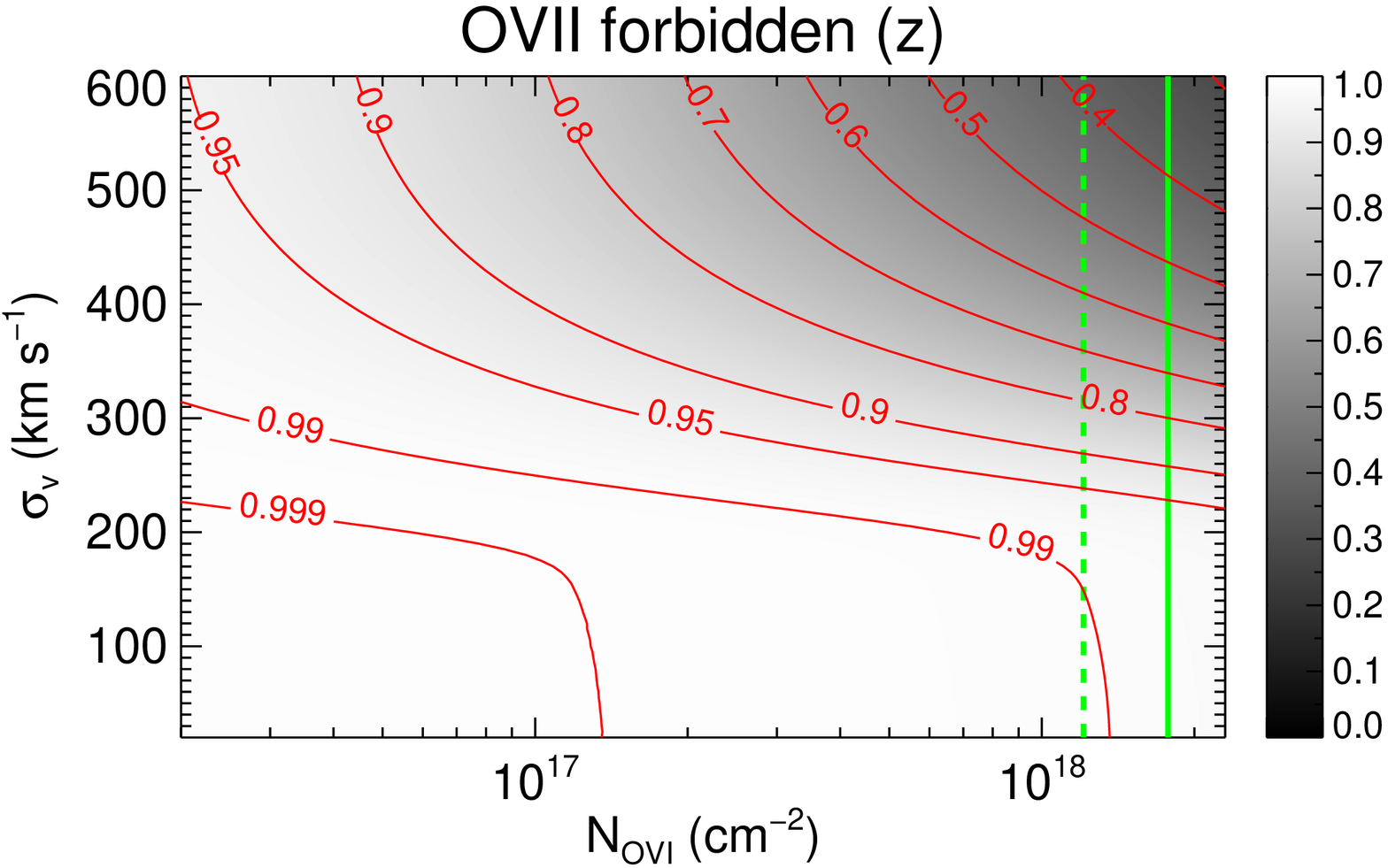}\hspace{1.0cm}\includegraphics[angle=0]{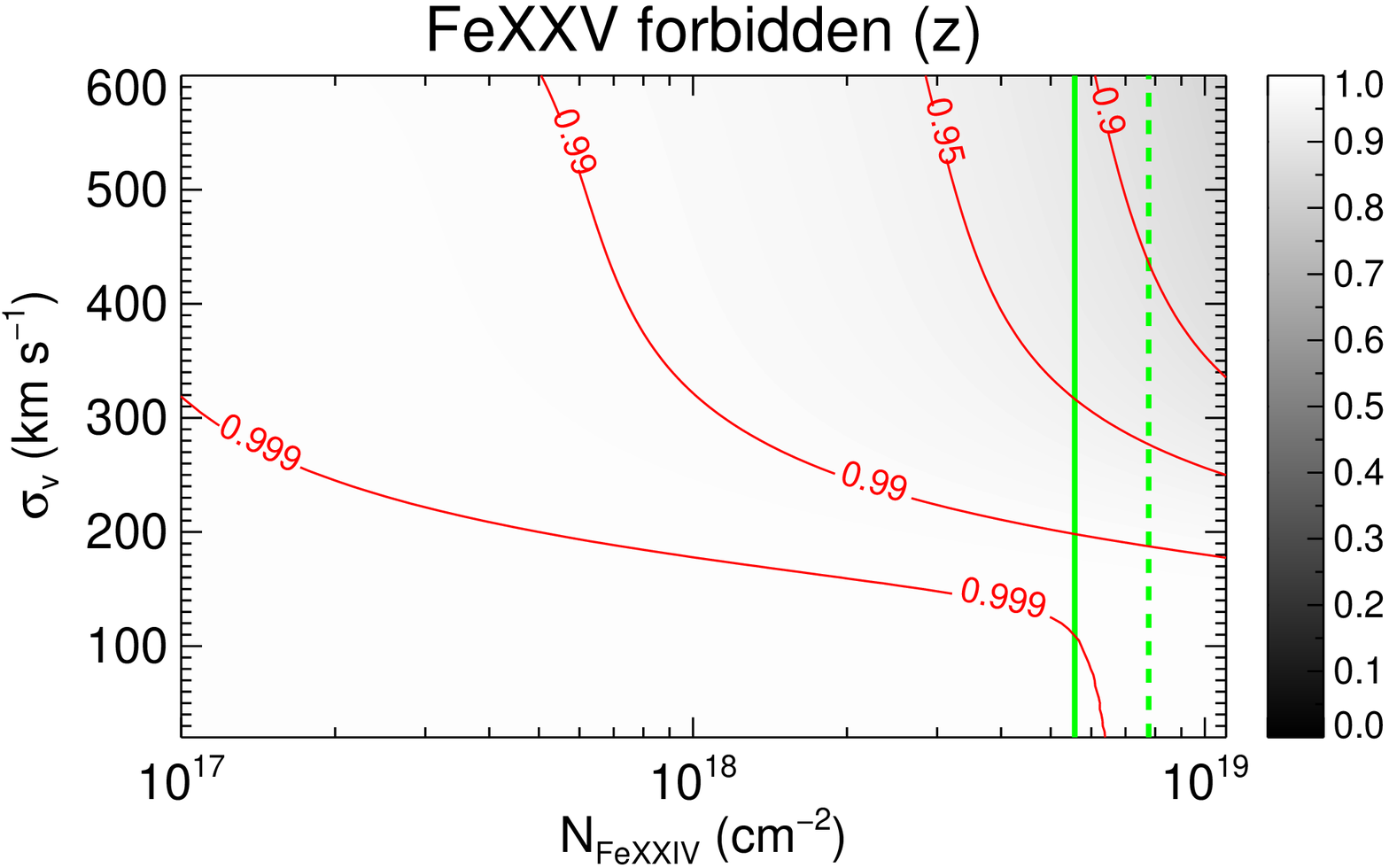}}
\caption{Contour diagrams showing the remaining flux fraction of the intercombination $y$ and forbidden $z$ emission lines of \ovii and \fexxv triplets, after intrinsic line absorption by the \ovi and \fexxiv ions (including fluorescent emission) as a function of $N_{\rm Li \mhyphen like}$ and $\sigma_{v}$. The green vertical lines in each panel indicate $N_{\rm crit}$, which is the critical Li-like ion column density limit corresponding to continuum optical depth ${\tau_{\rm cont} = 1}$ at the He-like triplet for CIE (dashed vertical line) and PIE (solid vertical line) plasma cases.}
\label{contour_fig}
\end{figure*}

Figure \ref{triplets_fig} shows examples of \ovii and \fexxv triplet emission lines, absorbed by their Li-like ion counterparts. The contour diagrams of Fig. \ref{contour_fig} show the emerging flux fraction of $y$ and $z$ lines of the He-like triplets after Li-like absorption and fluorescent emission as a function of $N_{\rm Li \mhyphen like}$ and \sigmav. The corresponding contour plot for the intercombination $x$ line is omitted to save space, as it looks identical (for \ovii) or similar (for \fexxv) to that of the $y$ line. In Fig. \ref{critical_fig} we show the column density limits at which continuum absorption and self-absorption of the He-like triplet lines (described in Sect. \ref{cont_abs_sect}) become critical in CIE and PIE plasmas. Finally, Fig. \ref{R_fig} shows how the unabsorbed $R$-ratio of the He-like triplets is altered because of \ovi and \fexxiv line absorption and fluorescent emission.

%

%

\subsection{Continuum absorption and self-absorption of the He-like ion triplet emission lines}
\label{cont_abs_sect}

\begin{figure}[!]
\centering
\resizebox{1.04\hsize}{!}{\hspace{-1.2cm}\includegraphics[angle=0]{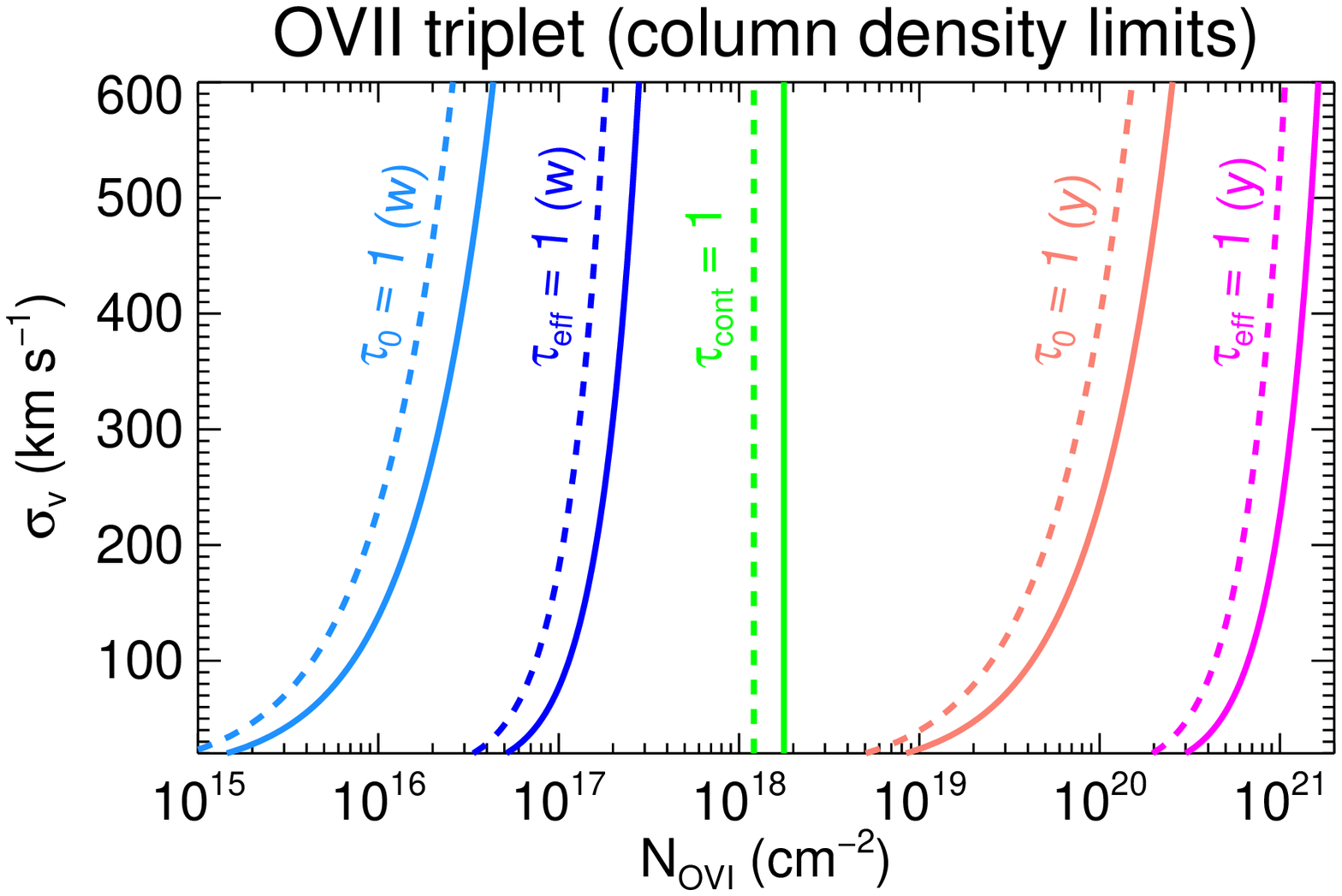}}
\resizebox{1.04\hsize}{!}{\hspace{-1.2cm}\includegraphics[angle=0]{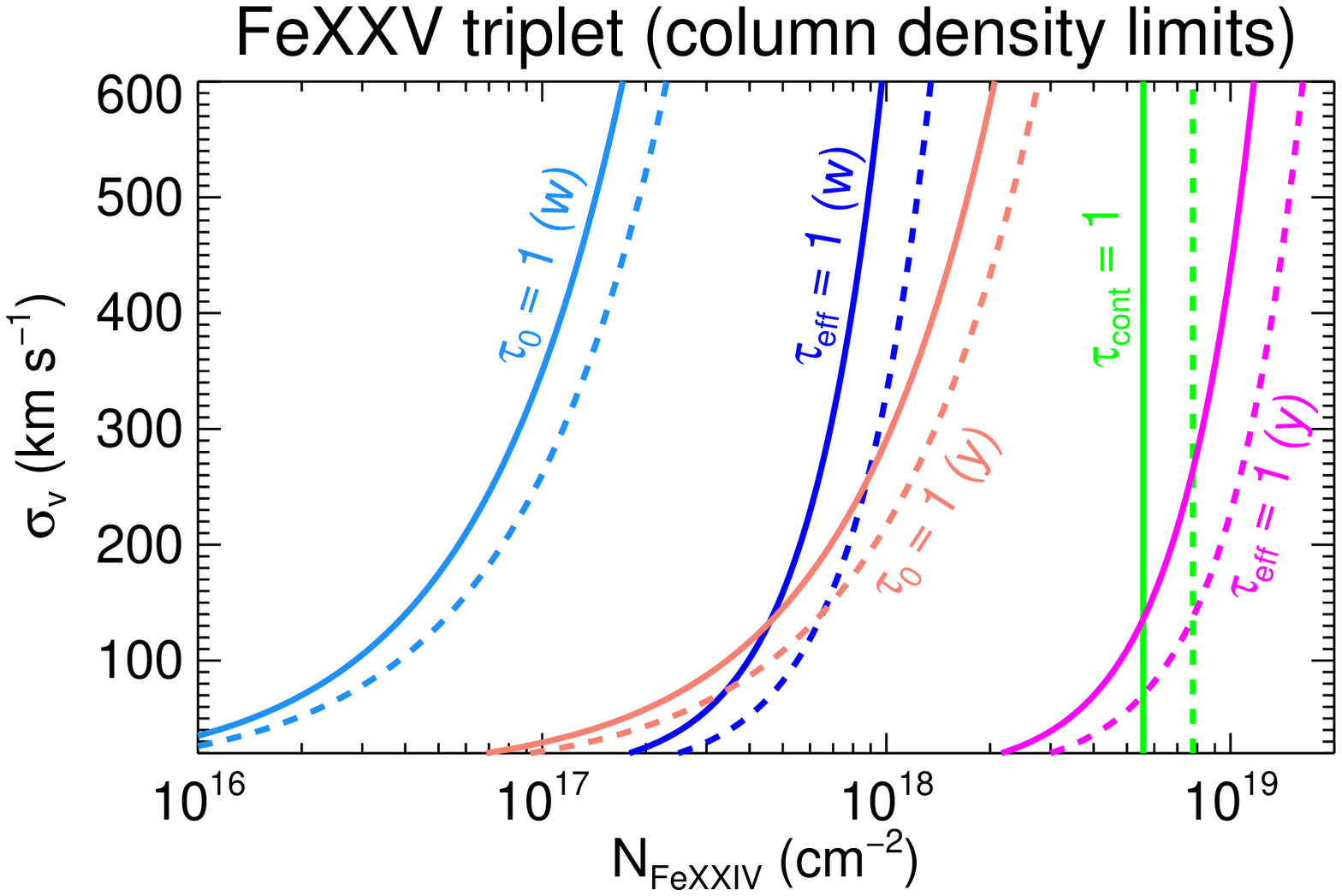}}
\caption{Contour diagrams corresponding to ${\tau_{0} = 1}$ and ${\tau_{\rm eff} = 1}$ limits for the resonance $w$ and intercombination $y$ line of \ovii ({\it top panel}) and \fexxv ({\it bottom panel}). The green vertical lines in each panel indicate $N_{\rm crit}$, which is the critical Li-like ion column density limit corresponding to continuum optical depth ${\tau_{\rm cont} = 1}$ at the He-like triplet. The dashed lines or curves are for the CIE plasma case, and the solid lines or curves are for the PIE plasma case as described in Sect. \ref{cont_abs_sect}.}
\label{critical_fig}
\end{figure}

\begin{figure}[!]
\centering
\resizebox{1.03\hsize}{!}{\hspace{-1.2cm}\includegraphics[angle=0]{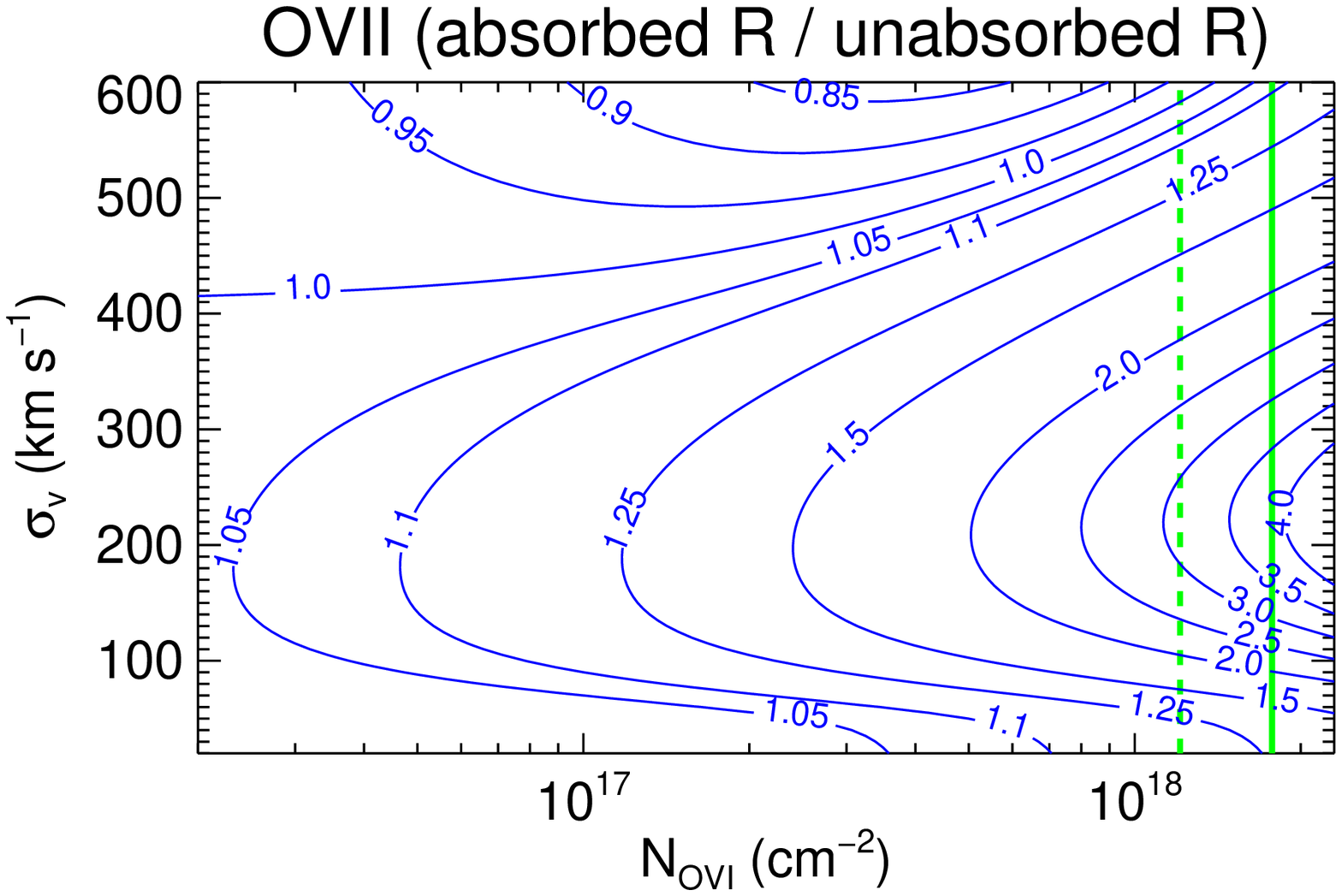}}
\resizebox{1.03\hsize}{!}{\hspace{-1.2cm}\includegraphics[angle=0]{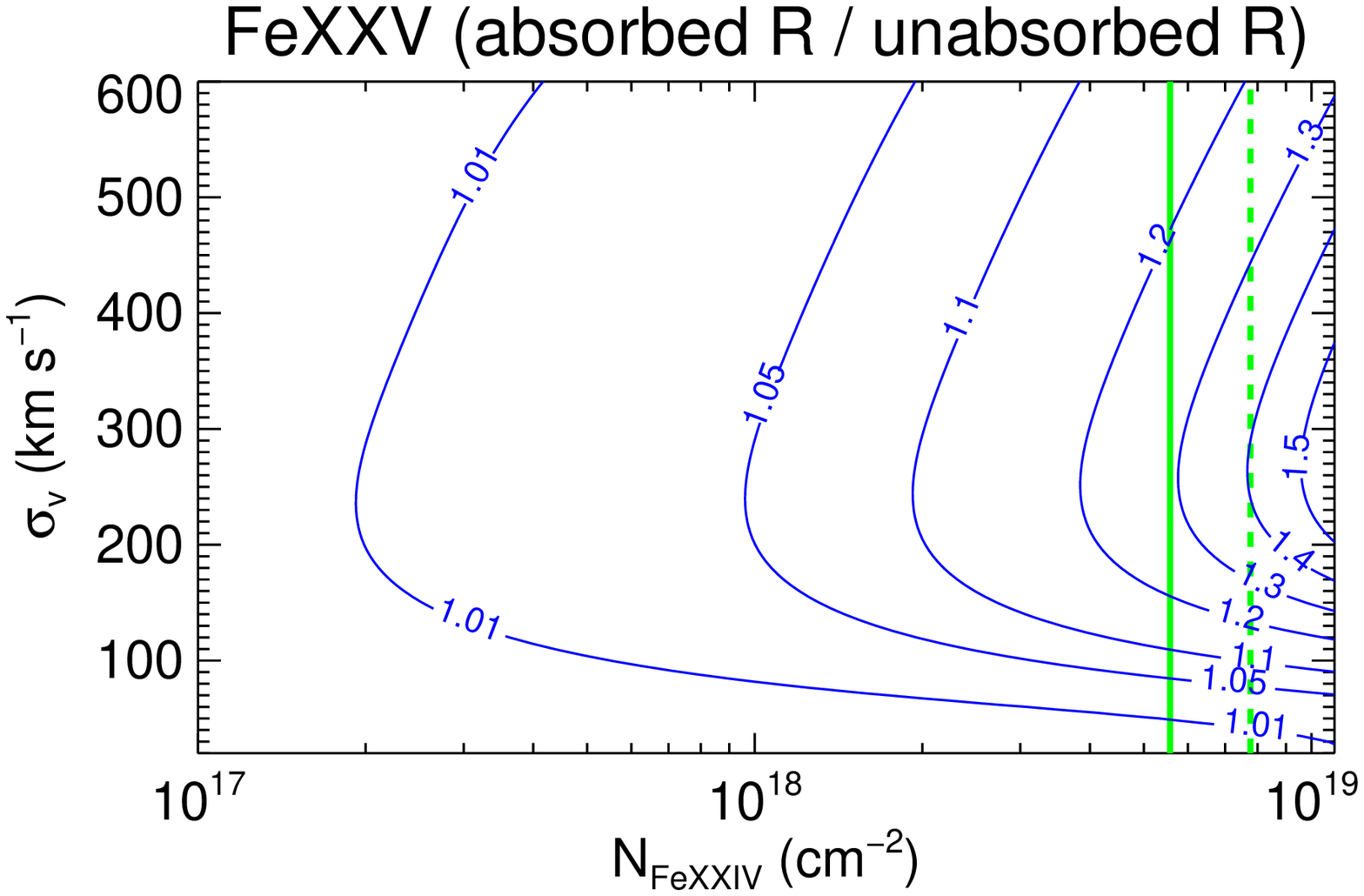}}
\caption{The absorbed $R$-ratio divided by the unabsorbed $R$-ratio for the \ovii ({\it top panel}) and \fexxv ({\it bottom panel}) triplets as a function of $N_{\rm Li \mhyphen like}$ and $\sigma_{\rm v}$. The contour lines show the factors by which the $R$-ratio value changes due to Li-like line absorption, including the effect of fluorescent emission. The green vertical lines in each panel indicate $N_{\rm crit}$, which is the critical Li-like ion column density limit corresponding to continuum optical depth ${\tau_{\rm cont} = 1}$ at the He-like triplet for PIE (solid vertical line) and CIE (dashed vertical line) plasma cases.}
\label{R_fig}
\end{figure}

At sufficiently high column densities, absorption of line photons in the continuum becomes important. For CIE and PIE plasma cases, we have computed the critical column density ($N_{\rm crit}$) of the Li-like ions for which the continuum optical ${\tau_{\rm cont} = 1}$ at the He-like ion triplet. We used the {\tt hot} (CIE) and {\tt pion} (PIE) models in the {\tt SPEX} code \citep{Kaa96} v2.05.04 to calculate $\tau_{\rm cont}$ for electron temperatures, where the ionic concentrations of \ovi and \fexxiv peak. For elemental abundances the proto-solar values of \citet{Lod09} were adopted. For the CIE case, these temperatures are found to be 0.0251 keV ($2.91 \times 10^5$ K) for \ovi and 1.58 keV ($1.83 \times 10^7$ K) for \fexxiv. For the PIE plasma case, we adopted the ionising Spectral Energy Distribution (SED) of \ngc obtained by \citet{Meh15a} (the 2013 `unobscured' SED version), which is representative of that of a typical Seyfert-1 AGN. In this PIE case, the temperatures at which the Li-like ionic concentrations peak are found to be 0.00177 keV ($2.05 \times 10^4$ K) for \ovi and 0.402 keV ($4.66 \times 10^6$ K) for \fexxiv, respectively. The $N_{\rm crit}$ limits (corresponding to ${\tau_{\rm cont} = 1}$ at the He-like triplets) for these CIE and PIE cases are overlaid on the contour diagrams of Figs. \ref{contour_fig}, \ref{critical_fig} and \ref{R_fig}. The \ovi $N_{\rm crit}$ is ${1.2 \times 10^{18}\ {\rm cm}^{-2}}$ for the CIE and ${1.8 \times 10^{18}\ {\rm cm}^{-2}}$ for the PIE case. The \fexxiv $N_{\rm crit}$ is ${7.8 \times 10^{18}\ {\rm cm}^{-2}}$ for the CIE and ${5.6 \times 10^{18}\ {\rm cm}^{-2}}$ for the PIE case.

In addition to continuum absorption, self-absorption of the He-like triplet lines also occurs. While the forbidden and intercombination line have in general a small optical depth $\tau_{0}$ for line absorption, the resonance line has a significant $\tau_{0}$. This is because $f_{\rm osc}$ of the resonance line is much higher than those of the intercombination and forbidden lines. For \ovii, $f_{\rm osc}$ is 0.696 for $w$, which is higher than $f_{\rm osc}$ of $x$, $y$ and $z$ by a factor of about ${5.86 \times 10^{6}}$, $5850$ and ${2.99 \times 10^{9}}$, respectively. For \fexxv, $f_{\rm osc}$ is 0.798 for $w$, which is higher than $f_{\rm osc}$ of $x$, $y$ and $z$ by a factor of about ${4.72 \times 10^{4}}$, $12.0$ and ${2.44 \times 10^{6}}$, respectively. Thus, after the $w$ line, the intercombination $y$ line has the highest $f_{\rm osc}$. The effective optical depth ($\tau_{\rm eff}$) for an absorption line includes contribution from both $\tau_{\rm cont}$ and $\tau_{0}$, which is given by ${\tau_{\rm eff} \approx \sqrt{\tau_{\rm cont} (\tau_{\rm cont} + \tau_{0})}}$ \citep{Ryb79}. Thus, if $\tau_{\rm eff} \ge 1$, the line has a significant probability of being absorbed on its scattering path through the medium. For the \ovii and \fexxv lines, we calculated $\tau_{0}$ and $\tau_{\rm eff}$ as a function of $N_{\rm Li \mhyphen like}$ and \sigmav. This is done for the above CIE and PIE plasma cases, for temperatures at which ionic concentrations of \ovi and \fexxiv peak. In Fig. \ref{critical_fig}, the contours corresponding to ${\tau_{0} =1}$ and ${\tau_{\rm eff} =1}$ for the resonance $w$ and the intercombination $y$ line of \ovii and \fexxv are shown. For the intercombination $x$ and forbidden $z$ line, the ${\tau_{0} =1}$ and ${\tau_{\rm eff} =1}$ limits (which are outside the displayed range of the plots in Fig. \ref{critical_fig}) are not relevant as they occur at very high column densities (${\gg N_{\rm crit}}$).

%

%

\section{Discussion and conclusions}
\label{discussion}

As evident in Figs. \ref{triplets_fig} and \ref{contour_fig}, Li-like ion line absorption can significantly diminish the flux of He-like ion triplet lines, in particular those of \ovii. The intercombination ($x$ and $y$) line is most affected by this line absorption. For instance, ${N_{\ovi} = 10^{18}\ {\rm cm}^{-2}}$, results in 35--80\% (depending on $\sigma_v$) of the emitted \ovii intercombination line to be observed (see Fig. \ref{contour_fig}). The forbidden $z$ line is much less affected by Li-like absorption than the intercombination line. The line absorption of $z$ is however more dependent on \sigmav than that of the intercombination line, becoming greater with increasing \sigmav. The resonance $w$ line remains almost completely unaffected by the Li-like line absorption (see Fig. \ref{triplets_fig}). The $w$ line is instead susceptible to absorption in the continuum due to its long path length caused by resonance scattering.

The predicted upper limit on the $R$-ratio for \ovii is about 4.0 for a PIE plasma according to {\tt Cloudy} calculations. This upper limit value decreases with the nuclear charge and reaches ${\sim 1}$ for \fexxv. For the examples shown in Fig. \ref{triplets_fig}, the $R$-ratio without Li-like absorption is 4.0 for \ovii and 1.0 for \fexxv. With Li-like absorption and the associated fluorescent emission, the emerging $R$-ratio changes from 4.0 to 9.7 for \ovii, and from 1.0 to 1.05 for \fexxv. Importantly, this is a large increase in the $R$-ratio for \ovii, whereas that of \fexxv increases by a small amount as \fexxiv line absorption of the \fexxv intercombination line is much less effective than \ovi line absorption of the \ovii intercombination line (see Fig. \ref{contour_fig}). This is due to significant fluorescent emission by \fexxiv, cancelling out the effect of its line absorption. Without fluorescent emission, the emerging $R$-ratio after absorption would be 9.5 for \ovii and 1.7 for \fexxv in the examples of Fig. \ref{triplets_fig}. Thus, the effect of fluorescent emission is small for \ovi absorption of the \ovii lines and much more significant for \fexxiv absorption of the \fexxv lines.

Figure \ref{contour_fig} shows that below the $N_{\rm crit}$ limit (i.e. in the optically-thin regime of ${\tau_{\rm cont} < 1}$), \ovi line absorption of the \ovii triplet intercombination line is significant. Thus, the resultant $R$-ratio will be significantly altered from its theoretical unabsorbed value. Figure \ref{R_fig} shows that \ovi line absorption causes the \ovii $R$-ratio to increase by a factor of up to 3.9 at ${\sigmav \approx 200\ \kms}$, hence the predicted upper limit of ${R \approx 4.0}$ for a PIE plasma can reach up to ${\approx 16}$ due to the \ovi line absorption. Indeed, this is consistent with the higher than expected \ovii $R$-ratio values found in AGN as described in Sect. \ref{triplet_obs_sect} (e.g. \citealt{Landt15}). Interestingly, Fig. \ref{R_fig} also reveals that at only high \sigmav (${\gtrsim 500}$~\kms), the $R$-ratio can actually become smaller in the optically-thin regime, due to higher line absorption of the forbidden line as \sigmav increases. For the \fexxv triplet, the maximum change in the $R$-ratio achieved by \fexxiv line absorption is only a factor of 1.3 in the PIE case and 1.4 in the CIE case.

We note that in both CIE and PIE plasmas, Li-like ion line absorption and hence $R$-ratio changes are possible in the optically-thin regime (see Fig. \ref{R_fig}). In principle, Li-like ion line absorption of the He-like ion triplet lines is intrinsic to any kind of plasma containing these ions. However, for this absorption to be able to significantly alter the $R$-ratio of the triplets and become observationally distinguishable, sufficiently high column density in our line of sight is required, such as those found in the PIE plasmas of AGN.

The effective optical depth ${\tau_{\rm eff}}$ of the resonance $w$ line (presented in Fig. \ref{critical_fig}) shows that ${\tau_{\rm eff} > 1}$ at relatively low column densities (i.e. below $N_{\rm crit}$). This means the $G$-ratio becomes altered as the $w$ line self-absorption becomes significant in the optically-thin regime. This is the case for both \ovii and \fexxv triplets in both CIE and PIE plasmas. On the other hand, ${\tau_{\rm eff}}$ of the intercombination and forbidden lines reach unity at much higher column densities (above $N_{\rm crit}$) due to their low $f_{\rm osc}$, and thus their impact on the $R$-ratio in the optically-thin regime is minimal. Although in the case of \fexxv intercombination $y$ line, the column density corresponding to ${\tau_{\rm eff}=1}$ starts below $N_{\rm crit}$ if ${\sigmav < 200}$~\kms. As long as ${N_{\rm Li \mhyphen like}}$ is lower than the column density corresponding to ${\tau_{\rm eff}=1}$ for the intercombination line (shown in Fig. \ref{critical_fig}), then even if ${N_{\rm Li \mhyphen like} > N_{\rm crit}}$, the $R$-ratio remains almost unchanged. This is because the wavelengths of the intercombination and forbidden lines are so close that they experience almost the same continuum absorption. 

We note that in the case of a transient plasma such as in supernova remnants (not investigated in the present study), the inner-shell ionisation of Li-like ions can also significantly increase the $R$-ratio value due to apparent increase in the intensity of the He-like forbidden line, in particular for high-$Z$ ions such as \fexxv. Additionally, the $R$-ratio is also dependent on the temperature of the plasma (albeit to lesser extent than the density) and reaches higher values in the case of a CIE plasma.

In this paper we use a simple slab geometry to demonstrate the principles of intrinsic Li-like line absorption and its effects on the He-like triplet emission lines without introducing additional complexities. We note that while for resonant scattering (hence changes in the $G$-ratio) the details of the plasma geometry become important, Li-like line absorption (and hence changes in the $R$-ratio) is primarily related to the optical depth of the medium in the line of sight, which is dependent on the column density and dispersion velocity as presented in this paper.

The future high-resolution X-ray spectrometers of the \astroh and {\it Athena} observatories will be useful for precise parameterisation of the He-like triplets lines. The measurements of the line parameters and flux using the existing grating spectrometers often give large uncertainties on the $R$-ratio due to low count-rate statistics. At present there are only very limited accurate measurements of the $R$-ratio from He-like triplets in the X-ray band (mostly \ovii) for photoionised plasmas. The microcalorimeter spectrometers of \astroh and {\it Athena} will combine high sensitivity with unprecedented spectral resolution at the 6~keV band. They will enable us to detect not only the He-like triplets in the soft X-ray band, but also the highly ionised \fexxv triplet at hard X-rays. This triplet is key for plasma diagnostics at the high-temperature and high-density domain. Using these upcoming instruments, the $R$-ratio and the associated Li-like line absorption can be investigated for different He-like triplets in a decent size sample of objects.

In conclusion, intrinsic line absorption by Li-like ions in a photoionised medium can significantly diminish the intensity of the intercombination line of its He-like ion triplet emitted inside the same medium. As a result, this absorption causes significant alteration in the line ratios of the triplets. This is the case for \ovi column densities between ${10^{17}}$ and ${10^{18}\ {\rm cm}^{-2}}$ in the optically-thin regime. The predicted $R$-ratio for \ovii, increases from 4 to an upper limit of 16 due to this process. For Li-like ions with higher nuclear charge (such as \fexxiv), the effect of line absorption becomes less apparent due to strong fluorescent emission by these ions. Finally, we emphasise that without considering the line absorption by Li-like ions, the use of observed He-like triplet line ratios can give erroneous density diagnostics for photoionised plasmas.

\begin{acknowledgements}

SRON is supported financially by NWO, the Netherlands Organization for Scientific Research. We thank the anonymous referee for her/his useful comments.

\end{acknowledgements}


\end{document}